# A connection between the structural α-relaxation and the β-relaxation found in bulk metallic glass-formers

K. L. Ngai[1*], Z. Wang[2], X. Q. Gao[2], H. B. Yu[2], W. H. Wang [2]

[1]*Dipartimento di Fisica, Università di Pisa, Largo B. Pontecorvo 3, I-56127, Pisa, Italy*

[2] *Institute of Physics, Chinese Academy of Sciences, Beijing 100190, People's Republic of China*

**Abstract**

New metallic glasses containing La or Ce have been introduced that have dynamic properties bordering on extremes of conventional metallic glasses. This provides opportunity to test trends or correlations established before in molecular and polymeric glass-formers if exist the same exist in the broader family of metallic glasses. Due to the drastically different chemical and physical structure of metallic glass-formers than soft matter, there is no guarantee that any correlation found in the latter will hold in the former. If found, the result brings metallic glasses closer to the much wider classes of glass-formers by the similarity in properties, and possibly have the same explanation. In non-metallic glass-formers, a general and fundamental connection has been established between the non-exponentiality parameter of the structural α-relaxation and the separation between its relaxation time $\tau_\alpha$ and the β-relaxation time $\tau_\beta$. In this paper we explore the experimental data of metallic glass-formers and show the correlation applies. An explanation of this correlation is given. The establishment of the correlation may facilitate the understanding of the roles played by the β-relaxation observed in macroscopic properties of metallic glasses including the relations to shear transformation zone, ductile-brittle transition in deformation, crystallization, and diffusion.

*Author to whom correspondence should be addressed: ngai@df.unipi.it

I. Introduction

Metallic glasses are unusual glass-forming materials in their structures made of multi-atomic particles in stark contrast to covalent bonded molecules in molecular and polymeric glassformers [1-6]. The interaction potential in metallic glass-formers [7] is also very different from the van der Waals interaction with or without the complication due to the presence of hydrogen bonds and ionic interaction. Naturally the following question arises. Beyond glass-forming ability, do metallic glasses share any of the outstanding and general dynamic property established before in much more abundant class of molecular and polymeric glasformers. Whether positive or negative, the answer would be interesting and could have far reaching implications.

In non-metallic glass-formers, there is a secondary relaxation with strong connection to the structural $\alpha$-relaxation manifested in various properties of dynamics [8-11]. This secondary relaxation involves the motion of the entire molecule or the entire monomer in the case of polymer. Its properties as well as its universal presence in all glass-formers are suggested by the analogous primitive relaxation of the Coupling Model. The single secondary relaxation found decades ago by Johari and Goldstein [12] in totally rigid molecular glass-formers such as chlorobenzene certainly involves the motion of the entire molecule. For this reason, the term Johari-Goldstein (JG) β-relaxation has been used to distinguish the secondary relaxation belonging to this important class from the other secondary relaxation of intramolecular nature and has no connection to the structural α-relaxation [8-11]. The multi-particle structure of the metallic glass-formers can pose a challenge to the universal presence of the JG β-relaxation for three reasons. Firstly, it is unclear if secondary relaxation exists in metallic glassformers. Secondly, even if it exists, its nature is unclear. Thirdly, it remains to be shown if the secondary relaxation exhibit any connection to the α-relaxation in any of the dynamic properties. Notwithstanding, secondary relaxation in different metallic glasses is either resolved [13] or unresolved but present as the excess loss [14] in isochronal and in isothermal mechanical relaxation measurements [15-17], and hence its existence seems assured. Recent synchrotron X-ray study of $Zr_{55}Cu_{30}Ni_5Al_{10}$ alloy has shown the secondary relaxation originates from a short-range collective rearrangement of atoms, implying that it involves motion of all atoms, and satisfying one major criterion of being the JG β-relaxation [9]. Still remain to be shown is any connection it may have to the α-relaxation in dynamic properties, before the conclusion can be made.

The purpose of the study reported in this paper is to provide one connection between the JG β-relaxation and the α-relaxation analogous to that found in non-metallic glass-formers. The link enables to make the conclusion that the secondary relaxation is the JG β-relaxation with fundamental importance as those of non-metallic glass-formers.

## 2. Correlation of $\log[\tau_\alpha(T_g)/\tau_\beta(T_g)]$ with $n(T_g)$: metallic glass-formers

To find out if there is correlation in metallic glass-formers, the two quantities, $\log[\tau_\alpha(T_g)/\tau_\beta(T_g)]$ and $n(T_g)$, have to be determined. This is not an easy task for any metallic glass-former. Although $T_g$ can be determined by calorimetry, and $\tau_\alpha(T_g)$ estimated from the scanning rate, the difficulty lies in acquiring $\tau_\beta(T_g)$ and $n(T_g)$. For molecular glass-formers several experimental techniques including dielectric relaxation, mechanical relaxation, light scattering and nuclear magnetic resonance are at disposal to obtain these two quantities, only mechanical relaxation is applicable to metallic glass-formers. The limited frequency range and lower sensitivity of mechanical relaxation make determination of $\tau_\beta(T)$ difficult at any temperature. Isothermal mechanical relaxation over many decades of frequencies are prerequisite for characterizing the structural α-relaxation and determining $n(T_g)$. At the present time, isothermal mechanical relaxation had been carried out on a few metallic glass-formers, and not all were able to resolve the β-relaxation. More customary are isochronal mechanical measurements, but $\tau_\beta(T_g)$ and $n(T_g)$ were obtained less directly with assumptions introduced to fit the temperature dependence of the isochronal mechanical loss data. Due to the aforementioned limitations of data of metallic glass-formers, larger errors accompanies of the values of $\tau_\beta(T_g)$ and $n(T_g)$ obtained from them. Therefore, if exists in metallic glass-formers, the correlation is not expected to be as precise as that found in molecular glass-formers [8-11,18]. This point has to be borne in mind when judging the correlation to be examined in the family of metallic glass-formers.

.

### A. *Experimental results from metallic glass-formers*

The isochronal shear loss modulus $G''$ measurement of the metallic glass-former, $Zr_{65}Al_{7.5}Cu_{27.5}$ at 5.4 kHz by Rösner et al. is perhaps the first that enabled deduction of $\tau_\beta(T_g)$ and $n(T_g)$ from the data [19]. From the Vogel-Fulcher-Tammann (VFT) temperature dependence of the α-relaxation time $\tau_\alpha$ of $Zr_{65}Al_{7.5}Cu_{27.5}$ (ZrAlCu) [14], values of $T_g$ defined operationally by either $\tau_\alpha(T_g)$=100 s or 1000 s were determined. The procedure used by Rösner et al. to extract the characteristics of the dynamics of the α-relaxation, which led to the $n(T_g)$=0.40, and log[$\tau_\beta(T_g)$/s]=-3.28 and -3.88 for $T_g$ defined $\tau_\alpha(T_g)$=100 s and 1000 s respectively [19]. Instead, we present the data of two recently studied metallic glass-formers, $La_{70}Ni_{15}Al_{15}$ (LaNiAl) and $Cu_{45}Zr_{45}Ag_{10}$ (CuZrAg), and the analyses that give estimates of $\tau_\beta(T_g)$ and $n(T_g)$ for them.

The DSC data of LaNiAl and CuZrAg at heating rate of 40 K/min are presented in Figs.1. From the data, $T_g$ of LaNiAl and CuZrAg are 441 and 684 K respectively. Isochronal loss modulus $E''$ data from 1 to 16 Hz of LaNiAl and CuZrAg are shown in Figs.2 and 3 respectively. The shifts of the lower temperature β-loss peak of LaNiAl and the excess loss of CuZrAg with frequency are give in the insets of Fig.2 and 3. From the shifts, the activation energy $E_\beta$ of the β-relaxation is estimated to be 90 kJ/mol for LaNiAl, and 205 kJ/mol for CuZrAg. The isochronal spectra at 1 Hz had been fitted globally with the contributions from the α- and β-relaxation assumed to be additive. The correlation function of the α-relaxation is assumed to be given by the Kohlrausch function, and the temperature dependence of $\tau_\alpha(T)$ has the Vogel-Fulcher-Tammann (VFT) form. The fits have been carried out before [20], and are reproduced in Figs.2 and 3. The VFT temperature dependence of $\tau_\alpha(T)$ and the Arrhenius T-dependence of $\tau_\beta(T)$ for $T<T_g$ have been determined by the fit. For LaNiAl, the global fit gives the activation energy $E_\beta$=87 kJ/mol slightly smaller than 90 kJ/mol obtained from the shift with frequency. The value of $E_\beta$=171 kJ/mol from the global fit for CuZrAg is significantly smaller than 205 kJ/mol from shift with frequency. The larger discrepancy of $E_\beta$ from the two methods is due to the fact that the β-relxation of CuZrAg is not resolved. All parameters relevant to test the correlation including $n(T_g)\equiv(1-\beta_{KWW})$ were determined from the global fits, and are listed in Table 1 for LaNiAl and CuZrAg. The values of $\beta_{KWW}$ are 0.42 and 0.62 for LaNiAl and CuZrAg respectively. The parameters for $Zr_{65}Al_{7.5}Cu_{27.5}$ (ZrAlCu) deduced from the data of Rösner et al. are also given there.

Isochronal shear loss modulus $G''$ of $Ce_{70}Al_{10}Cu_{20}$ was measured at the frequencies of 0.5, 1, 2 and 5 Hz by Liu et al. [16]. The low temperature loss show no feature that can be identified with the presence of the β-relaxation. Essentially the same fitting procedure as describe in the above was used to fit the exclusively the α-loss peak with the combination of the Kohlrausch function for the time dependence and VFT equation for $\tau_\alpha(T)$ [16]. The value of $\beta_{KWW}$=0.8 was obtained from the fit. In addition to isochronal measurements, Liu et al. made isothermal shear modulus measurements. Their data at several temperatures together with the fits by the Fourier transform of the Kohlrausch function are reproduced in Fig.4. Remarkably the same value of $\beta_{KWW}$=0.8 fits the isothermal data. The excess wing of $G''(f)$ is evident from the deviation of the experimental data over the Kohlrasuch fit. The location of the arrows indicate the primitive rrealxation frequencies at 343 and 353 K calculated from the Coupling Model equation to be discussed in the subsection (2) later.

There are a few other metallic glass-formers which had been studied by isothermal mechanical relaxation, and $n(T_g)$ determined directly by the fit to the Fourier transform of the Kohlrausch stretched exponential function [15,21,22] . However not all of them have complementary data to determine $\tau_\beta(T)$ for $T<T_g$, and hence they cannot be used to test the correlation. The only cases we can find that can be used for our purpose are $Zr_{46.75}Ti_{8.25}Cu_{7.5}Ni_{10}Be_{27.5}$ (Vit4) [15], and $Pd_{40}Ni_{10}Cu_{30}P_{20}$ [21]. The value of $\tau_\beta(T_g)$ is obtained by extrapolating the experimentally determined Arrhenius $T$-dependence of $\tau_\beta(T)$ for $T<T_g$ to $T_g$. The relevant parameters, $\tau_\beta(T_g)$, $n(T_g)$, and $T_g$ defined by $\tau_\alpha(T_g)=10^3$ s of these metallic glass-formers are entered into Table 1.

Putting the results obtained for the six metallic glass-formers altogether in a plot of of $\tau_\beta(T_g)$ versus $n(T_g)$, the correlation between these two quantities can be tested. The plot is presented in Fig.4. The data of $Ce_{70}Al_{10}Cu_{20}$ are not sufficient to determine $\tau_\beta(T_g)$, and the point representing it in Fig.4 is the primitive relaxation time, $\tau_0(T_g)$, of the CM to be introduced in the next section.

Table 1. The values of the relevant parameters, $\tau_\beta(T_g)$, $n(T_g)$, and $T_g$ defined by $\tau_\alpha(T_g)=10^3$ s of the six metallic glass-formers considered.

| Metallic glass-former | $T_g$ (K) | log[ $\tau_\beta(T_g)$ /s] | $n(T_g)$ | $E_a/RT_g$ |
|---|---|---|---|---|
| $Zr_{65}Al_{7.5}Cu_{27.5}$ [14,19] | 606 | -3.28 | 0.40 | N/A |
| $Cu_{45}Zr_{45}Ag_{10}$ [20], and this study | 684 | -1.93 | 0.38 | 30.1±5.3 |
| $Zr_{46.75}Ti_{8.25}Cu_{7.5}Ni_{10}Be_{27.5}$ [15] | 620 | -3.3 | 0.50 | 26.8 |
| $Pd_{40}Ni_{10}Cu_{30}P_{20}$ [21] | 593 | -3.64 <br> -3.64 | 0.43 [21] <br> 0.50 [22] | 26.1 |
| $Ce_{70}Al_{10}Cu_{20}$ [16] | 343 | N/A | 0.2 | N/A |
| $La_{70}Ni_{15}Al_{15}$ [20], | 441 | -4.7 | 0.58 | 23.8 |

| | and this study | | | | |

N/A: not available.

The shear mechanical relaxation data of three other La metallic glass-formers, $La_{68.5}Ni_{16}Al_{14}Co_{1.5}$, $La_{60}Ni_{15}Al_{25}$, and $La_{57.5}Ni_{12.5}Al_{17.5}Cu_{12.5}$, and $Zr_{55}Cu_{30}Ni_5Al_{10}$ published by Qiao et al. [12] are helpful to show the correlation. It can be seen from the normalized frequency dependence of the loss modulus $G''(f/f_{max})/G_{max}$ that higher the La content in the metallic glass-former, the larger the full-width at half-maximum of the α-loss peak, and larger is the separation between the β- and the α-loss peaks or equivalently between $\log\tau_\alpha$ and $\log\tau_\beta$.

## B. Explanation from CM, and generality

Conceptually, the primitive relaxation of the Coupling Model (CM) and the JG β-relaxation are equivalent in the sense that both are independent relaxation acting as the precursor of the structural α-relaxation. Their properties are similar, especially those demonstrating the strong connection to the α-relaxaiton. Furthermore such as pressure dependence. Furthermore, $\tau_{JG}$ is approximately the same as the primitive relaxation time $\tau_0$ at any temperature $T$ and pressure $P$ in the supercooled liquid state of many glass-formers [8-11,18], i.e.,

$$\tau_\beta(T,P) \approx \tau_0(T,P) \tag{1}$$

where $\tau_0$ is calculated from the parameters, $\tau_\alpha$ and $n$, of the *α*-relaxation via the relation,

$$\tau_\alpha = [t_c^{-n}\tau_0]^{\frac{1}{1-n}}. \tag{2}$$

In Eq.(2) $t_c$ is the temperature insensitive crossover time from primitive relaxation with time dependence, $\exp(-t/\tau_0)$, to cooperative α-relaxation with the Kohlrasuch stretched exponential time dependence,

$$\varphi(t) = \exp\left[-(t/\tau_\alpha)^{1-n}\right] \tag{3}$$

The fraction of unity, (1-n), in Eq.(2) is the same as the stretch exponent $\beta_{KWW} \equiv (1-n)$ in Eq.(3). In context of the CM, $n$ is the coupling parameter. The value of $t_c$ is about 1 to 2 ps for soft molecular glass-formers and polymers as determined by quasielastic neutron scattering experiments and molecular dynamics simulations [11]. It is shorter for metallic systems with $t_c \approx 0.2$ ps from simulations [18,19], which is reasonable as inferred from the stronger metallic bonds compared with van der Waals interaction in soft matter, and is also exemplified by the much higher shear modulus of BMG than molecular glasses [25,26].

From the CM equations (2) and (3), the separation between $\tau_\alpha$ and $\tau_\beta$ (all quantities in seconds) given by,

$$\log_{10}\tau_\alpha - \log_{10}\tau_\beta = n(\log_{10}\tau_\alpha - \log_{10}t_c), \tag{4}$$

is proportional to $n$. This explains the empirical relation between the separation and $n$ found generally in molecular and polymeric glass-formers, and could also applicable to the six metallic glass-formers in Table 1. At $T=T_g$ defined by $\tau_\alpha(T_g)=10^3$ s, Eq.(4) has the simplified form of

$$\log_{10}\tau_\beta(T_g) = (1-n)\log_{10}\tau_\alpha(T_g) + n\log_{10}t_c = 3(1-n) - 12.7n, \tag{5}$$

which is represented by the red line in Fig.6. The trend in the dependence of $\tau_\beta(T_g)$ on $n(T_g)$ is in accord with that from Eq.(5). However, $\tau_\beta(T_g)$ of three of the six metallic glass-formers (LaNiAl, CuZrAg and Vit4) are shorter than that predicted value by about one and half decade. Discrepancy of this order of magnitude is not unexpected because of the uncertainty introduced to obtain $\tau_\beta(T_g)$ on $n(T_g)$ in LaNiAl and CuZrAg. Taking the possible errors into consideration, we can conclude that the metallic glass-formers also obey the correlation of $\log[\tau_\alpha(T_g)/\tau_\beta(T_g)]$ with $n(T_g)$ as found generally in molecular, polymeric and even inorganic glass-formers [6].

### C. Correlation of $\log[\tau_\alpha(T_g)/\tau_\beta(T_g)]$ with $m(T_g)$

While the frequency dispersion and hence $n(T_g)$ of metallic glass-formers is hard to obtain, more accessible from experiments is the fragility or steepness index, $m = d\log\tau_\alpha(T_g/T)/d(T_g/T)$ evaluated at $T_g/T=1$. On the other hand, by plotting the isochronal mechanical loss against $T/T_g$, a qualitative measure of the separation between $\tau_\alpha$ and $\tau_\beta$ can be deduced. This was done by Yu et al. [27] in their study of the family of $(Ce_xLa_{1-x})_{68}Al_{10}Cu_{20}Co_2$ for $0 \leq x \leq 1$, They found Ce-richer member has smaller value of $m$ and the $\beta$-relaxation is closer to the $\alpha$-relaxation and shows up as an excess wing. La-richer member has larger $m$ and its $\beta$-relaxation is more separated from the $\alpha$-relaxation and becomes resolved either as a shoulder or peak. In this manner, Yu et al. found correlation between fragility $m$ and a property of the β-relaxation, namely the degree of separation between $\tau_\alpha$ and $\tau_\beta$ in the metallic glass-formers. However one should be mindful that this very correlation breaks down in some molecular glassformers. A well known example is propylene carbonate which has a large $m=104$ [28] but the β-relaxation is too close to the α-relaxation and appears as an excess wing [29]. In contrast, the correlation between $\log[\tau_\alpha(T_g)/\tau_\beta(T_g)]$ with $n(T_g)$ remains valid.

It is true that when restricting the glass-formers considered to a special class, correlation between properties with $m$ holds, as found by Yu et al. [27] in the family of (Ce$_x$La$_{1-}$ $_x)_{68}Al_{10}Cu_{20}Co_2$ for $0 \leq x \leq 1$. Notwithstanding, it remains a nontrivial task to derive any observed correlation with $m$. Here we provide a derivation of the correlation between $\log[\tau_\alpha(T_g)/\tau_\beta(T_g)]$ with $m$ found in metallic glass-formers. Established from the studies of many different metallic glass-formers is that in the glassy state the activation energy of β-relaxation is related to $T_g$ by $E_\beta = 26(\pm 2)RT_g$. [30]. Hence, in a plot against $T_g$-scaled reciprocal temperature, $T_g/T$, $\log_{10}\tau_\beta(T_g/T)$ has the same slope of 26/2.303=11.3 (but not necessarily the same magnitude) for all metallic glass-formers with different values of $n(T_g)$ or Kohlrausch exponent $\beta_{KWW}(T_g) \equiv (1- n(T_g))$. The fact that the variation of $\log_{10}\tau_\beta(T_g/T)$ with $T_g/T$ for $T_g/T>1$ is the same for all the metallic glasses considered is remarkable, and it implies that this holds also for $T_g/T<1$ as well. This is shown From Eq.(1) Again applying Eqs.(1) and (2) together, we have for all $T_g/T$ the relation

$$\tau_\alpha\left(\frac{T_g}{T}\right) = \left[(t_c)^{-n}\tau_0\left(\frac{T_g}{T}\right)\right]^{\frac{1}{1-n}} \approx \left[(t_c)^{-n}\tau_\beta\left(\frac{T_g}{T}\right)\right]^{\frac{1}{1-n}} \tag{6}$$

To satisfy this relation, the dependence of the magnitude of $\tau_0(T_g/T)$ or $\tau_\beta(T_g/T)$ as well as $\tau_\alpha(T_g/T)$ on the metallic glass-former through $n$ becomes clear. It is easy to see from Eq.(6) that at $T_g/T=1$ where $\tau_\alpha(T_g/T)=10^3$ s, $\log_{10}\tau_0(T_g/T)$ or $\log_{10}\tau_\beta(T_g/T) = [3/(1-n) - (12.7)n]$. This dependence of $\log_{10}\tau_0(T_g/T)$ on $n$ at $T_g/T=1$ is demonstrated for $n$=0.2, 0.3, 0.4 and 0.5 in Fig.7, and also the parallel shifts of $\log_{10}\tau_0(T_g/T)$ with $T_g/T$ for $T_g/T>1$ and $T_g/T<1$. In the final step, we calculate $\tau_\alpha(T_g/T)$ for all the four chosen values of $n$.

The results shown in Fig.7 clearly demonstrate larger $n$ engenders steeper rise of $\tau_\alpha(T_g/T)$ or more fragile behavior. The fragility index $m$ have been calculated, and the values are 25, 31.3, 35.7, 41.7, and 50 respectively for $n$=0.0, 0.2, 0.3, 0.4, and 05. Thus the correlation between fragility index *m* and the degree of separation between $\tau_\alpha$ and $\tau_\beta$ observed experimentally in metallic glass-formers [27] is now derived from the CM.

## 3. Conclusion

Some universal properties of the secondary β-relaxation of the Johari-Goldstein kind have been well established in soft glass-formers including molecular and polymeric substances. Among them is the relation of the secondary relaxation to the structural α-relaxation determined by the fractional exponent, $\beta_{KWW}\equiv(1-n)$, of the Kohlrausch-Williams-Watts correlation function used to characterize the frequency dispersion of the α-relaxation. Specifically, the separation between $\tau_\alpha$ and $\tau_\beta$ measured by $[\log\tau_\alpha(T_g) - \log\tau_\beta(T_g)]$ at $T=T_g$ not only correlates with $n(T_g)$ as found empirically by experiments, but also can be explained quantitatively by the Coupling Model. Metallic glass-former differs drastically in chemical and physical structures than soft matter. Composed of a collection of atoms, it is not obvious that the β-relaxation exists in metallic glass-former. Interestingly, β-relaxation exists in metallic glasses, and are important from the connections it has to shear transformation zones, ductile-brittle transition in deformation, crystallization, and diffusion.

Although the β-relaxation exists in metallic glass-formers, it is not clear if the correlation will hold. Despite the difficulty in obtaining both quantities, $\tau_\beta(T_g)$ and $n(T_g)$, in metallic glass-formers, we have found $[\log\tau_\alpha(T_g) - \log\tau_\beta(T_g)]$ correlates with $n(T_g)$ at $T=T_g$. Taking into consideration the uncertainties of these parameters deduced by fitting experimental data, the correlation found is quantitatively consistent with the CM prediction. The establishment of the correlation may facilitate the understanding of the roles played by the β-relaxation observed in macroscopic properties of metallic glasses.

**Acknowledgment**

The work performed at the Institute of Physics, Chinese Academy of Sciences, Beijing was supported by the NSF of China (51271195and11274353).

**References**

[1] A. L. Greer, Science **267**, 1947 (1995).

[2] W. Dmowski, T. Iwashita, C. P. Chuang, J. Almer, and T. Egami, Phys. Rev. Lett. **105**, 205502 (2010).

[3] T. Egami, Prog. Mater. Sci. **56**, 637 (2011).

[4] H. Wagner, D. Bedorf, S. Küchemann, M. Schwabe, B. Zhang, W. Arnold, and K. Samwer, Nature Mater. **10**, 439 (2011).

[5] Y. H. Liu, D. Wang, K. Nakajima, W. Zhang, A. Hirata, T. Nishi, A. Inoue, and M. W. Chen, Phys. Rev. Lett. **106**, 125504 (2011).

[6] A. R. Yavari, K. Georgarakis, W. J. Botta, A. Inoue, and G. Vaughan, Phys.Rev.B **82**, 172202 (2010).

[7] D.B. Miracle, Nature Mater. **3**, 697 (2004).

[8] K. L. Ngai, J. Chem. Phys. **109**, 6982 (1998).

[9] K. L. Ngai and M. Paluch, J. Chem. Phys. **120**, 857 (2004).

[10] S. Capaccioli, M. Paluch, D. Prevosto, Li-Min Wang, and K. L. Ngai, J. Phys. Chem. Lett. **3**, 735 (2012).

[11] K. L. Ngai, *Relaxation and Diffusion in Complex Systems* (Springer, New York, 2011).

[12] . G. P. Johari, and M. Goldstein, J. Chem. Phys. **53**, 2372 (1970).

[13] J.M. Pelletier, B. Van de Moortele, and I.R. Lu, Materials Science and Engineering, **A336**, 190 (2002).

[14] P. Rösner, K. Samwer, and P. Lunkenheimer P., Euro.Phys.Lett. **68,** 226 (2004).

[15] P. Wen, D.Q. Zhao, M.X. Pan, W.H. Wang, Y.P. Huang, and M.L. Guo, Appl. Phys. Lett. **84**, 2790 (2004).

[16] X.F. Liu, B. Zhang, P. Wen, W.H. Wang, J. Non-Cryst. Solids **352**, 4013 (2006).

[17] J. C. Qiao and J. M. Pelletier, J. Appl. Phys. **112**, 083528 (2012).

[18] S. Capaccioli, K. Kessairi, D. Prevosto, M. Lucchesi and P. A. Rolla, J. Phys.: Condens. Matter **19**, 205133 (2007).

[19] K.L. Ngai, **352**, 404 (2006).

[20] Z. Wang, P. Wen, L. S. Huo, H. Y. Bai, and W. H. Wang, Appl. Phys. Lett. **101**, 121906 (2012).

[21] Z. F. Zhao, P. Wen, C. H. Shek, and W. H. Wang, Phys. Rev.B **75**, 174201 (2007).

[22] Li-Min Wang, Riping Liu, and Wei Hua Wang, J. Chem. Phys. **128**, 164503 (2008).

[23] H. Teichler, Phys. Rev. Lett. **76**, 62 (1996).

[24] X. J. Han and H. R. Schober, Phys.Rev.B **83**, 224201 (2011).

[25] Mingwei Chen, Annu. Rev. Mater. Res. **38**, 445 (2008).

[26] Wei Hua Wang, Prog. Mater. Sci. **57**, 487 (2012).

[27] H.B. Yu, Z. Wang, W.H. Wang, H.Y. Bai, J. Non-Cryst. Solids, **358** (2012) 869.

[28] R. Böhmer, K.L. Ngai, C.A. Angell, D.J. Plazek, J. Chem. Phys. **99**, 4201 (1993).

[29] K.L. Ngai, P. Lunkenheimer, C. León, U. Schneider, R. Brand, A. Loidl, J. Chem. Phys.**115**, 1405 (2001).

[30] H. B. Yu, W. H. Wang, H. Y. Bai, Y. Wu, and M. W. Chen, Phys. Rev. B

**81**, 220201 (2010).

**Figure Captions**

Fig. 1. The DSC curve of $La_{70}Ni_{15}Al_{15}$ (red curve) and $Cu_{45}Zr_{45}Ag_{10}$ (blue curve) measured at 40 K/min.

Fig. 2. Temperature dependence of the loss modulus for a $La_{70}Ni_{15}Al_{15}$ MG, measured with frequency from 1 to 16 Hz, at the heating rate of 3 K/min.

Fig. 3. Temperature dependence of the loss modulus for a $Cu_{45}Zr_{45}Ag_{10}$ MG, measured with frequency from 1 to 16 Hz, at the heating rate of 3 K/min.

Fig. 4. Temperature dependence of loss modulus of $La_{70}Ni_{15}Al_{15}$ measured at 1 Hz. The colored regions represent the KWW fit. The inset is a plot of $\log f$ against $1000/T_p$, where $T_p$ is the peak temperature of the β-relaxation.

Fig. 5. Temperature dependence of loss modulus of $Cu_{45}Zr_{45}Ag_{10}$ measured at 1 Hz. The colored regions represent the KWW fit. The inset is a plot of $\log f$ against $1000/T_p$, where $T_p$ is the peak temperature of the β-relaxation.

Fig. 6. The frequency dependence of the normalized loss modulus of $Ce_{70}Al_{10}Cu_{20}$ determined at temperatures of 343, 353, 363 and 373 K (from left to right). The solid lines are the fits by the combination of the KWW function with $\beta_{KWW} \equiv (1-n) = 0.80$, and the VFT equation. The arrows indicate the logarithm of the primitive frequencies, $\log f_o$, calculated from the CM equation.

Fig.7. Plot of log[$\tau_\beta(T_g)$] at $T=T_g$ against $\beta_{KWW} \equiv (1-n)$, of the Kohlrausch-Williams-Watts correlation function used to characterize the frequency dispersion of the $\alpha$-relaxation. La stands for $La_{70}Ni_{15}Al_{15}$, Cu stands for $Cu_{45}Zr_{45}Ag_{10}$, Ce stands for $Ce_{70}Al_{10}Cu_{20}$, and Pd stands for $Pd_{40}Ni_{10}Cu_{30}P_{20}$. The two points for Pd correspond to the two values of $\beta_{KWW}$ given in the literature by different authors. For more details, see text.

Fig.8. The parallel lines for $T_g/T>1$ from top to bottom represent the log[$\tau_\beta(T)$] of metallic glasses with increasing values, 0.2, 0.3, 0.4, and 0.5, of the coupling parameters $n \equiv (1-\beta_{KWW})$. Here $T_g$ is defined by log[$\tau_\alpha(T_g)$]=3. The location of log[$\tau_\beta(T_g)$] at $T=T_g$ for each value of $n$ is determined by Eq.(6). The dependence of $\tau_\beta$ on $T_g/T$ for $T_g/T<1$ is taken to be the same for all $n$, as suggested by the case for $T_g/T>1$. The lines for $T_g/T<1$ are log[$\tau_\alpha(T_g/T)$] calculated from from $\tau_\beta(T_g/T)$ by Eq.(6) for the values of $n$=0.0, 0.2, 0.3, 0.4, and 0.5. The same color and style of the lines are used for $\tau_\alpha(T_g/T)$ and $\tau_\beta(T_g/T)$ for each $n$ value.

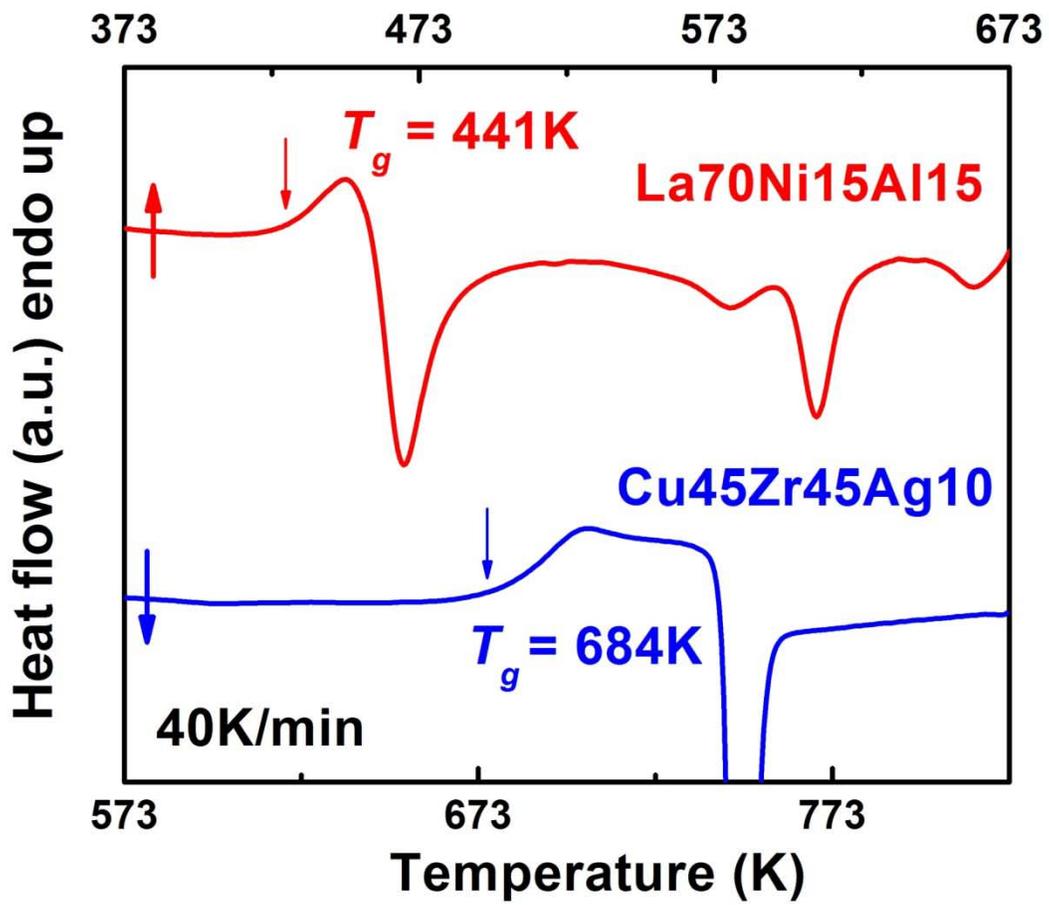

Fig.1

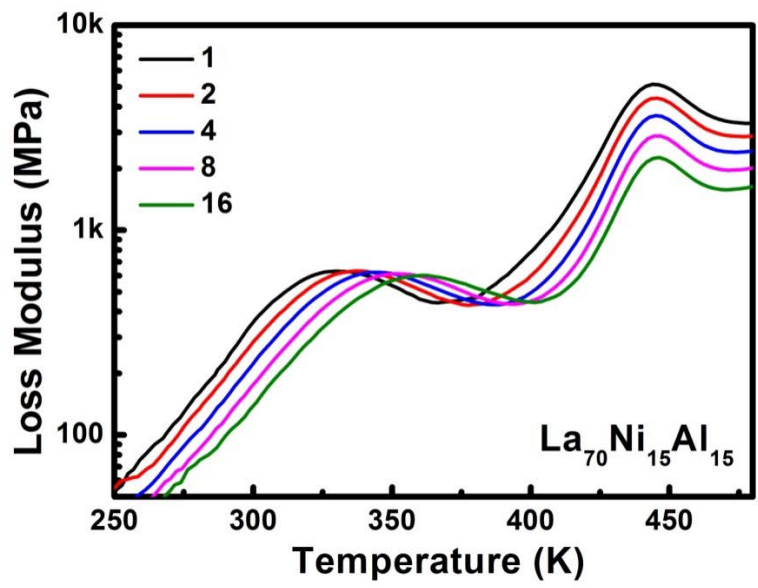

Fig.2

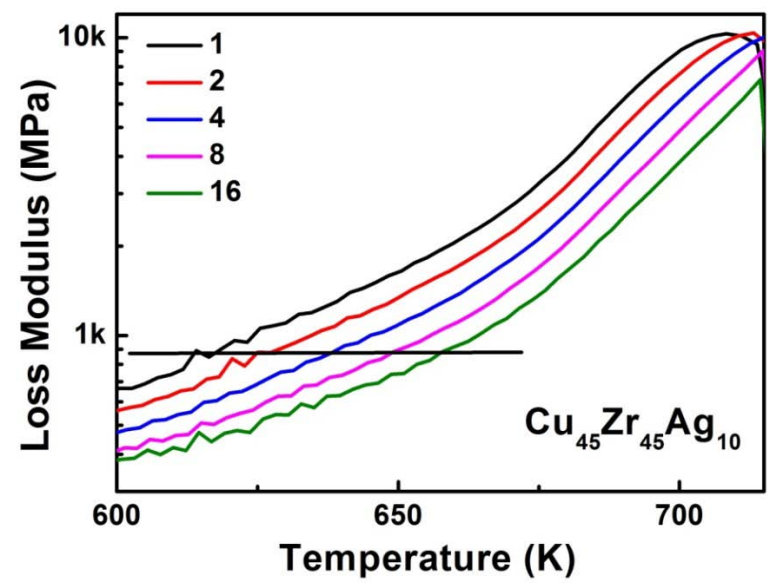

Fig.3
18

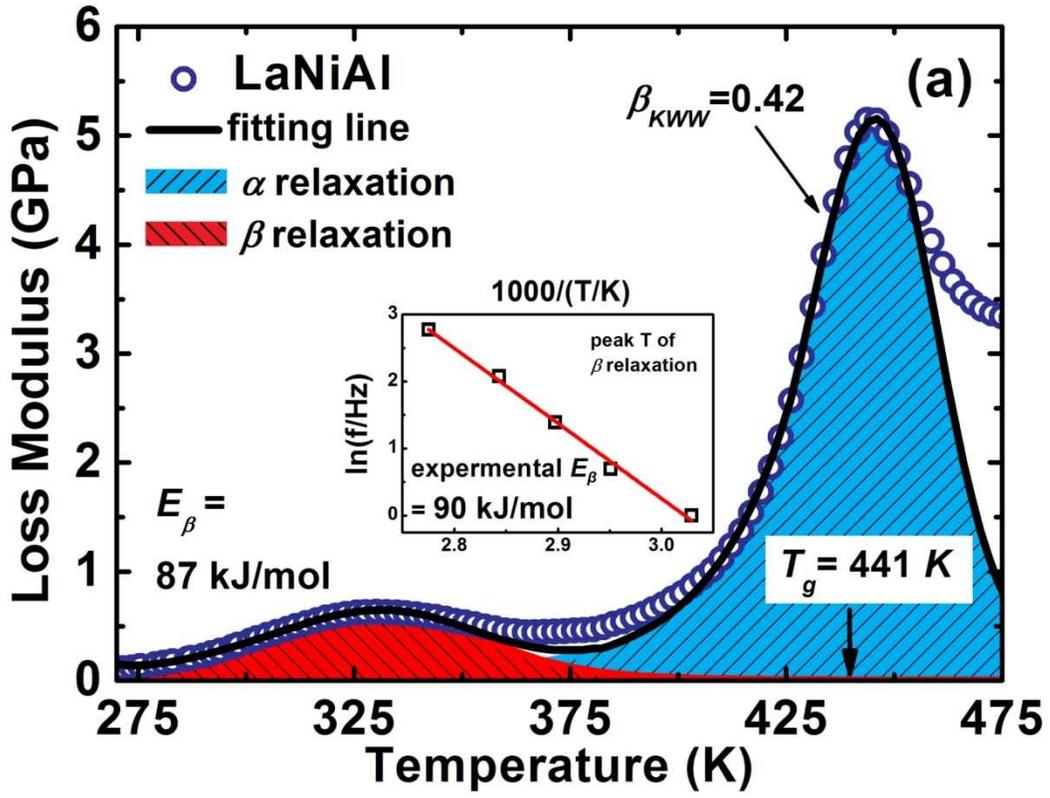

Fig.4

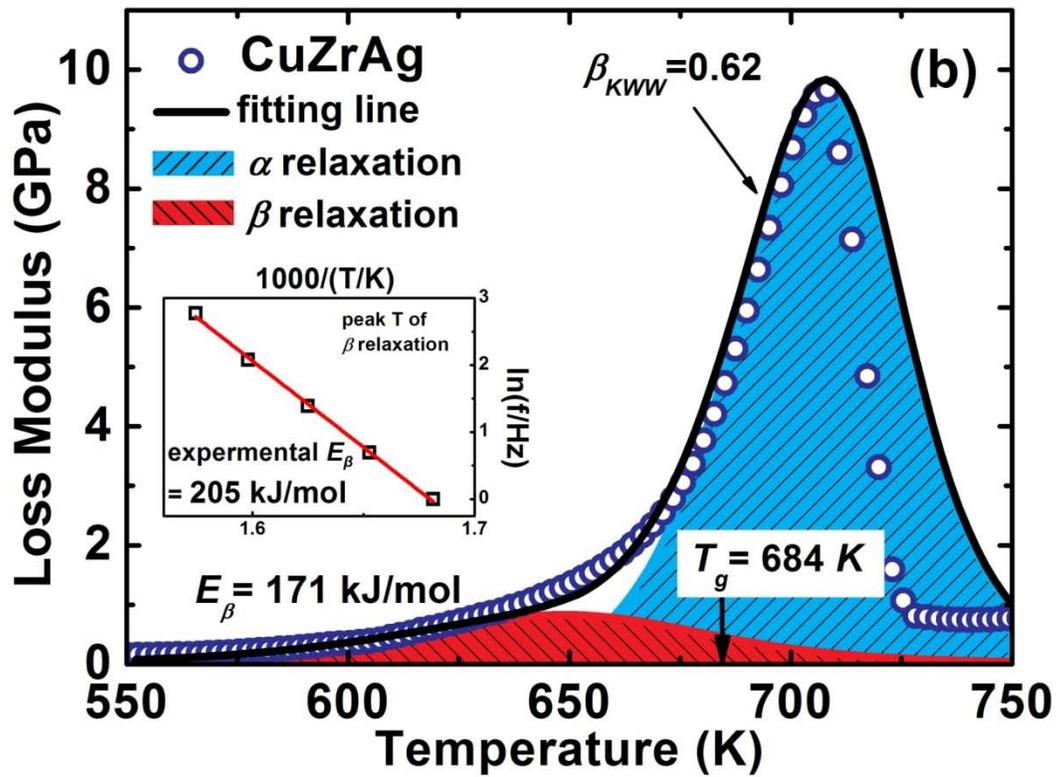

Fig.5

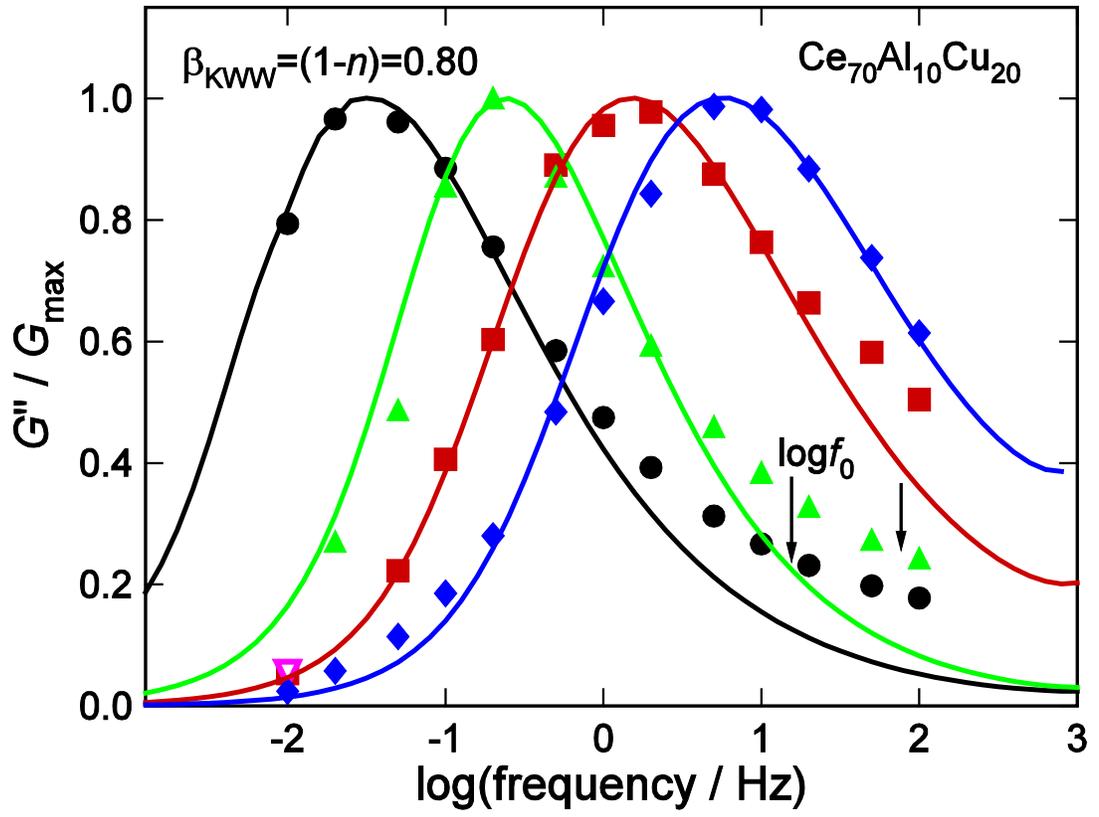

Fig.6.

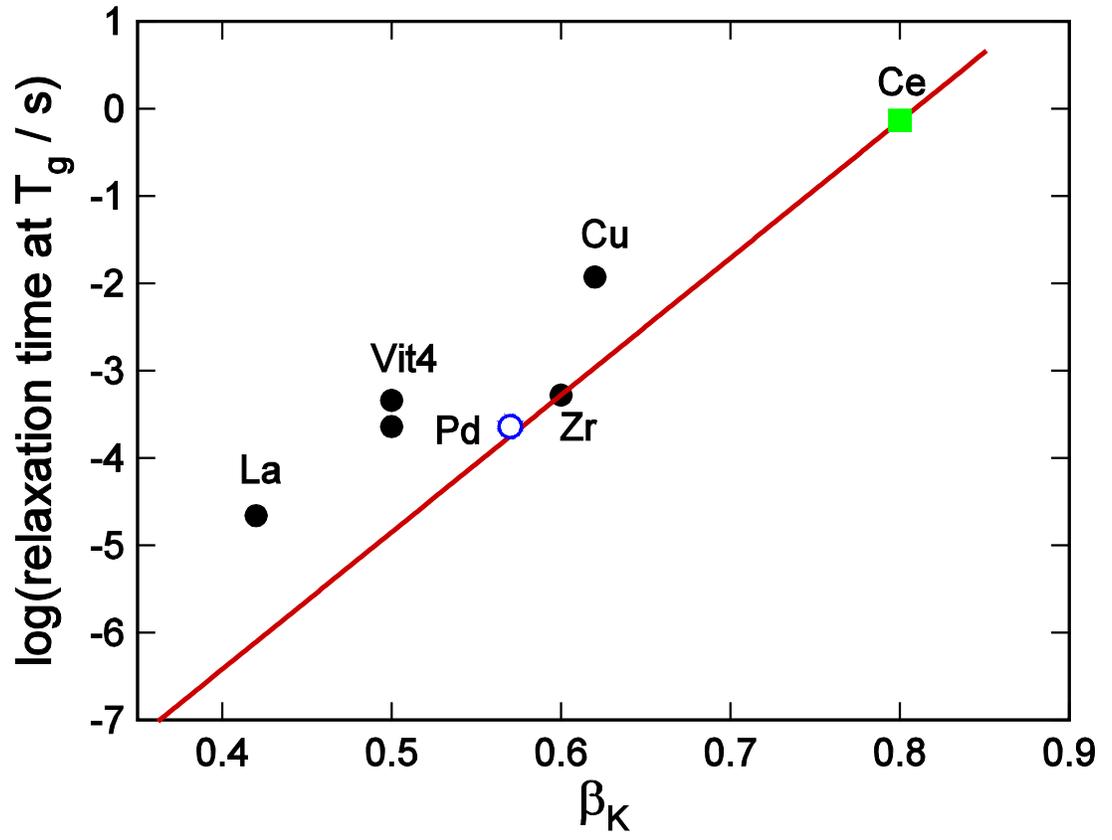

Fig.7.

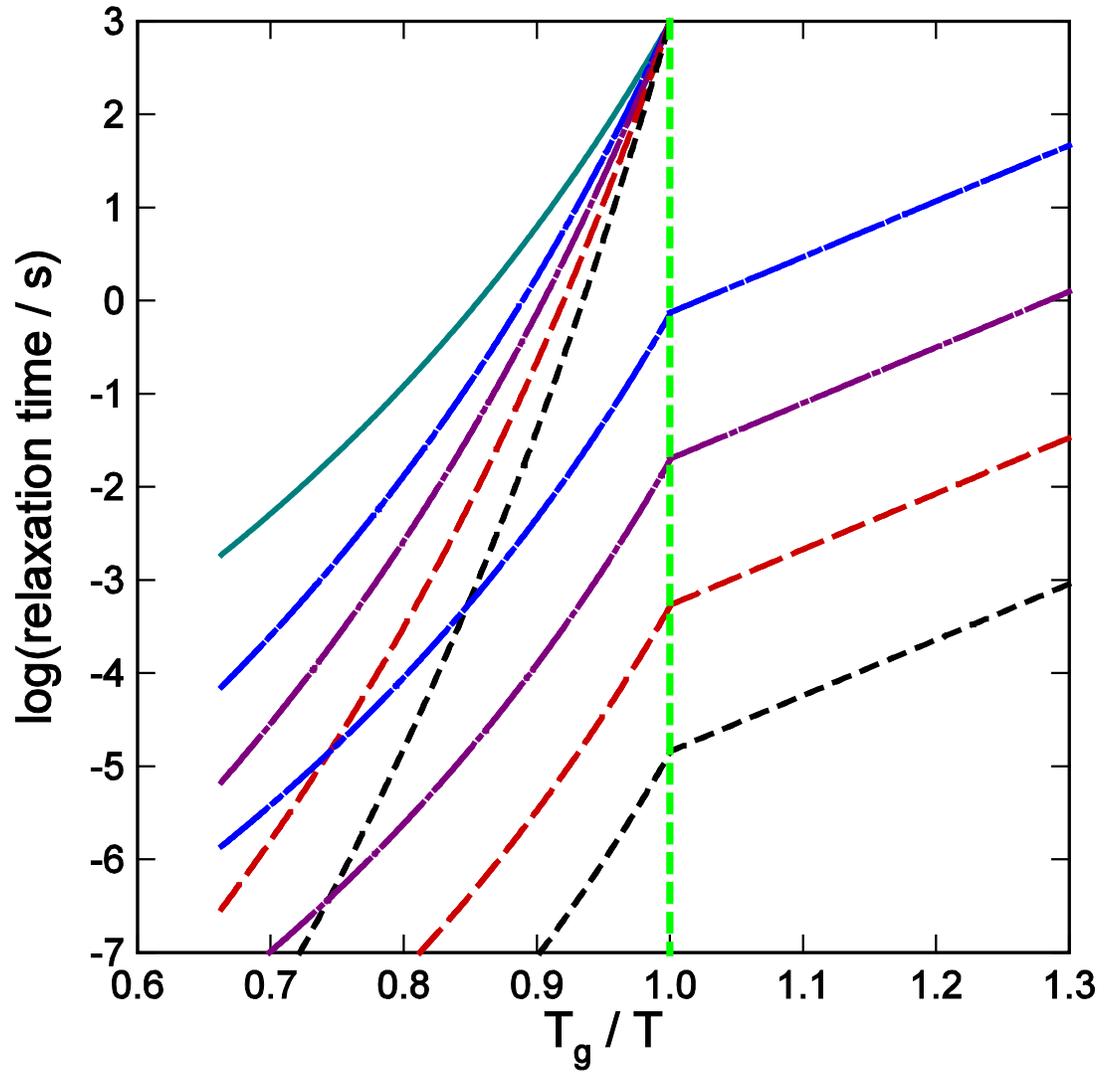

Fig.8